\documentclass{jetpl}
\twocolumn

\lat

\title{Investigation of LiFeAs by means of "Break-junction" Technique}

\rtitle{LiFeAs Superconducting Gaps Determination by the "Break-junction" Technique \ldots}

\sodtitle{LiFeAs Superconducting Gaps Determination by the "Break-junction" Technique}

\author{S.\,A.\, Kuzmichev$^{1}$\/\thanks{e-mail: kuzmichev@mig.phys.msu.ru}, T.\,E.\,Shanygina$^{1,2}$, I.\,V.\,Morozov$^1$, A.\,I.\,Boltalin$^1$, M.\,V.\, Roslova$^1$, S.\,Wurmehl$^3$, B.\,B\"{u}chner$^3$}

\rauthor{S.\,A.\, Kuzmichev, T.\,E.\,Shanygina, I.\,V.\,Morozov, \emph{et al.}}

\sodauthor{S.\,A.\, Kuzmichev, T.\,E.\,Shanygina, I.\,V.\,Morozov, \emph{et al.}}

\address{$^1$M.\,V.\,Lomonosov Moscow State University,
119991 Moscow, Russia\\~\\
$^2$P.\,N.\,Lebedev Physical Institute of the RAS,
119991 Moscow, Russia\\~\\
$^3$IFW-Dresden, Institute for Solid State Research, D-01171 Dresden, Germany}

\abstract{In our tunneling investigation using Andreev superconductor - normal metal - superconductor contacts on LiFeAs single crystals we observed two reproducible independent subharmonic gap structures at dynamic conductance characteristics. From these results, we can derive the energy of the large superconducting gap  $\Delta_L=(2.5 \div 3.4)$ meV and the small gap  $\Delta_S=(0.9 \div 1)$ meV at $T = 4.2$ K for the $T_C^{local} \approx (10.5 \div 14)$ K (the contact area critical temperature which deviation causes the variation of $\Delta_L$). The BCS-ratio is found to be $2\Delta_L/k_BT_C = (4.6 \div 5.6)$, whereas $2\Delta_S/k_BT_C \ll 3.52$ results from induced superconductivity in the bands with the small gap.}


\begin{document}

\maketitle

The new class of superconducting rare-earth oxypnictides \cite{Kamihara} is still not completely understood and therefore requires further investigation. The layered LiFeAs (111-system) \cite{Wang} is one of the few stoichiometric Fe-based pnictides which shows neither magnetic nor structural transition but becomes superconducting at 18 K \cite{Borisenko,Heyer}. Band-structure calculations \cite{Singh,Nekrasov,Coldea} show the Fermi surfaces for 111-system to be comprised of quasi-two-dimensional (2D) hole cylinders centered at the $\Gamma$-point and electron ones at the M-point of the first Brillouin zone that can be considered as two effective bands (so called minimal two-band model) \cite{Raghu,Li}. The total density of states at the Fermi level is formed mainly by Fe 3d-states \cite{Kurmaev,Miyake,Platt,Jishi,Nekrasov2}. As was shown in \cite{Kuchinskii}, the superconducting transition temperatures $T_C$ for different types of iron-based superconductors correlate with the total density of states at the Fermi level. This fact and the strong Fe isotope effect, which was reported by \cite{Liu} supports the phonon-mediated coupling importance \cite{Bardeen} in these compounds \cite{Kuchinskii,Boeri}. The electron-phonon coupling is enhanced by an extended van Hove singularity \cite{Abrikosov} which was shown for iron pnictides and, in particular, for LiFeAs \cite{Kordyuk}. In this work, we present an investigation of the superconducting properties of LiFeAs single crystals by means of Andreev spectroscopy of superconductor - normal metal - superconductor (SNS) contacts, and the corresponding superconducting gaps.

The LiFeAs single crystals were obtained by self-flux method. The synthesis and investigation of the composition and properties are detailed in \cite{Morozov}. A mixture of small lumps of the Li metal and powders of Fe and As in a molar ratio of Li : Fe : As = 3 : 2 : 3 was placed into an alumina crucible. All work on the reactive mixture preparation was carried out in a dry box under argon atmosphere. The crucible was inserted into a niobium container, which was welded in argon at 1.5 atm. The sealed Nb container was enclosed in a quartz ampoule. The sample was heated up to 1363 K, kept at this temperature and slowly cooled down to 873 K. At this temperature ampoule was extracted from the furnace and cooled in open air. The LiFeAs single crystals in the form of thin plates with lateral dimensions of $(12 \pm 6)  \times  (12 \pm 6)  \times  (0.1 \pm 0.05)$ mm$^3$ were separated from flux mechanically. According to the XRD, EDX, ICP MS, and nuclear quadruple resonance spectroscopy (NQR) obtained crystals have a stoichiometric composition with a homogeneous distribution of elements over the entire sample \cite{Morozov}.

Due to the strong hygroscopicity of LiFeAs, the rectangular plates were kept in glass capillaries. The "break-junction" technique \cite{Moreland,Muller} was exploited to make SNS-Andreev contacts. The sample mounting on a spring holder was made in argon atmosphere in order to prevent the material from decomposition in air. The samples (thin plates with dimensions about $(2 \times 1 \times  0.2)$ mm$^3$) were attached to the holder with two current and two potential leads by liquid In-Ga alloy. In order to measure on cryogenically clean surfaces, the holder was bended by a micrometric screw at $T = 4.2$ K causing a microcrack in the sample. The relatively small separation distance between two superconducting banks prevents any extrinsic impurity penetration into the microcrack. The current-voltage curves $I(V)$, their derivatives $dI(V)/dV$ and $R(T)$-dependences were measured by installation controlled by AT-MIO-16X (National Instruments) digital board \cite{LOFA,Rakhmanina}. The dynamic conductance spectra ($dI(V)/dV$) were obtained following a standard current modulation technique.

Resistive measurements showed the superconducting transition of our Li$_{1-\delta}$FeAs single crystals at $T^{bulk}_C$ ranging from 11\,K to 17\,K (see Fig.1), which may be caused by a minor affecting of water vapors and leads to a slight variation of lithium amount. As was shown theoretically in \cite{Shein}, while lithium deficiency turns Li$_{1-\delta}$FeAs system into antiferromagnetic state (non superconducting) at $\delta=0.5$, structure of the Fe-As layers is not changed and $c$ lattice parameter is decreased by $\sim14\%$ \cite{Shein}. Obviously, the samples with critical temperature reduced up to $\sim0.7\,T_C^{max}$ (e.g. filled circles at Fig.1) are described by a small structure deviation from the stoichiometric one. To the best of our knowledge, a variety of physical measurements on the single crystals obtained from the same batch were carried out in \cite{Borisenko,Heyer,Kordyuk,Morozov, Stockert,Inosov}, see also Table 1 concerning results on $T_C$ and gap values. The samples with the maximal $T^{bulk}_C \approx 17$\,K (see open circles at Fig.1, as well as data published in \cite{Morozov}) demonstrated a relatively narrow superconducting transition $(1.2 \div 1.5)$\,K, which confirms high quality of single crystals.

To evaluate the superconducting gaps two related methods were used: Andreev spectroscopy \cite{Andreev} of individual superconductor-constriction-superconductor (ScS) \cite{Kummel} Sharvin-type contacts \cite{Sharvin} and intrinsic Andreev spectroscopy (intrinsic multiple Andreev reflections effect that usually exists due to steps-and-terraces presence at clean cryogenic clefts). Unlike SN-Andreev contacts, symmetrical ScS-contacts simplify the interpretation of the dynamic conductance spectra by using theoretical model by K\"{u}mmel et al. \cite{Kummel}. The main current-voltage characteristic (CVC) features of our ScS-contacts involve an excess current at low voltages and a subharmonic gap structure (SGS), showing series of dips of a dynamic conductance $dI/dV$ at certain bias voltages \cite{Kummel,Aminov}:
\begin{equation}
V_n = \frac{2\Delta}{en},~~n=1,\,2\dots
\end{equation}
due to the multiple Andreev reflections effect. In a case of a multi-gap superconductor several independent SGSs corresponding to the each gap should be observable (see Figs.2-5). Such two distinct SGSs were reported earlier at Andreev spectra of Mg(Al)B$_2$ break-junctions \cite{SSC04,JETPL04,SSC12}, LaO$_{0.9}$F$_{0.1}$FeAs \cite{LOFA}, GdO(F)FeAs \cite{Gd,UFN,GdJPCS}, and FeSe \cite{FeSe}. The main advantage of a symmetrical SNS-contact is that at any $T < T_C$ the gap energy can be defined directly from the bias voltages of the peculiarities at the $dI/dV$-curve making use of the formula (1). The "break-junction" technique allows one to use such the advantages. The experimental CVCs obtained in this work are typical for the clean classical SNS-contact with excess-current characteristic \cite{Kummel,Aminov}, therefore, the model by K\"{u}mmel et al. \cite{Kummel} is valid to describe our data. Strictly speaking, the sharpest SGS usually presents only at $dI(V)/dV$-characteristic of the most qualitative Andreev contacts with a small diameter $a$ which is less than mean free path of the quasiparticles $l$ (ballistic limit) \cite{Sharvin}. In such a case one can observe several gap peculiarities, that increases the accuracy of the gap energies definition. Please note that if $a \approx l$, only the $n = 1$ minima would make a valuable contribution to the dynamic conductance spectra, which was the case for some LiFeAs Andreev contacts studied.

\begin{figure}[hc]
\includegraphics[width=.45\textwidth]{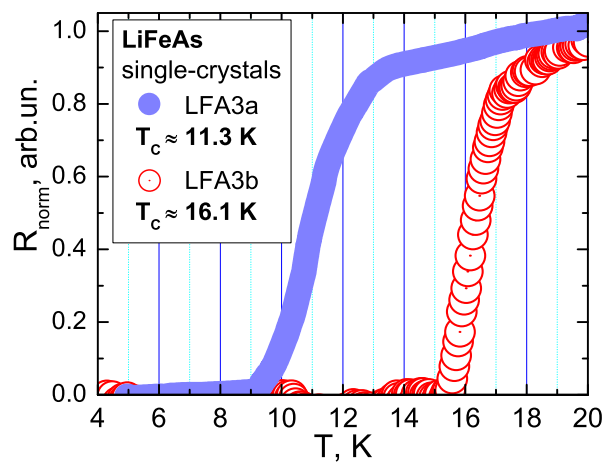}
\caption{Fig.1. Superconducting transitions for LiFeAs single crystals measured before a microcrack forming. The $dR(T)/dT$-curve maxima were assigned as $T_C^{bulk}$. The bulk critical temperatures range from 11 K to 17 K}
\end{figure}

A significant structure anisotropy of LiFeAs allows us to observe stack contacts \cite{Nakamura,Matsui} representing a sequence of SNS-junctions and intrinsic multiple Andreev reflections effect, which was observed earlier by the "break-junction" technique in other layered superconductors (Bi-2201 \cite{Bi}, Mg(Al)B$_2$ \cite{SSC04,JETPL04,SSC12}, and some Fe-based superconductors \cite{LOFA,FeSe,GdJPCS} along the $c$-direction). The bias voltage at $dI(V)/dV$-characteristic of such an array scales with the number of contacts $N$ (in a stack) in comparison with dynamic conductance of an individual contact. Existence of natural stacks in c-direction on cryogenic clefts makes usage of such structures typical for "break-junction" technique. The array contacts provide reducing of surface defects influence (which otherwise significantly affect the properties of superconductor \cite{Heumen}) by $N$ times. Therefore, the magnitude of bulk gaps can be measured with a higher accuracy using stack contacts.

Fig.2 represents the excess-current CVC and the corresponding derivatives for two SNS-Andreev break-junctions (3 contact arrays normalized to a single junction; were formed by a mechanical readjustment) in LiFeAs single crystal measured at $T = 4.2$ K. The main features at the dynamic conductance curve are located at $V_1 = 2\Delta_ L/e \approx  5$ mV. The origin of the fine structure is not yet clear and requires further investigation. The observed subharmonic gap structure (SGS) dips $n_L = (1, 2, 3)$ are seemingly related to the large gap (marked by black labels). Peculiarities at $V  \approx 3.5$ mV may appear due to an anisotropy of the $\Delta_L$ order parameter. However, we need further studies to confirm this unambiguously. The SGS bias voltages described by formula (1) (see inset of the Fig.2) can easily yield the value  $\Delta_L \approx 2.5$ meV (at the $T_C^{local} \approx 10.5$ K) without any additional fittings. The "local" $T_C$ is referred to as the $T_C$ measured locally at the contact point with diameter $a \approx l$ and might differ from the $T_C$ obtained by non-local methods. It is worth to note that location of Andreev reflexes were not moved under a mechanical readjustment of the contact while the contact resistance changed (see $I_{7\sim}(V)$ and $I_8(V)$ at Fig.2). Thus, one can conclude that the gap values are still unchanged. This means high homogeneity of the sample superconducting properties in the measuring range.

\begin{figure}[hc]
\includegraphics[width=.45\textwidth]{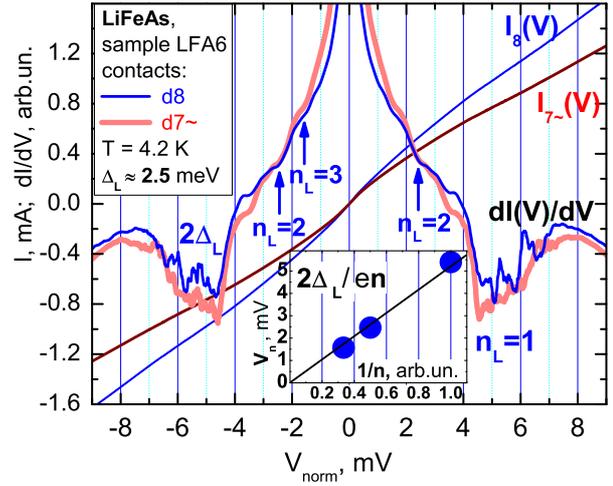}
\caption{Fig.2. Normalized to a single junction CVC and dynamic conductance of Andreev arrays $\sharp d7^{\sim}$ (bold light curve) and $\sharp d8$ (dark thin line) formed by 3 SNS-junctions in a stack ($T = 4.2$ K). The subharmonic gap structure (SGS) minima ($n_L$ labels) define the large superconducting gap $\Delta_L\approx 2.5$ meV which is well-reproduced after the mechanical readjustment of the contact (the $T_C^{local} \approx 10.5$ K). Inset: the linear SGS-minima bias voltages $V_{nL}(1/n)$-dependence for the aforementioned characteristics}
\end{figure}
\begin{figure}[hc]
\includegraphics[width=.455\textwidth]{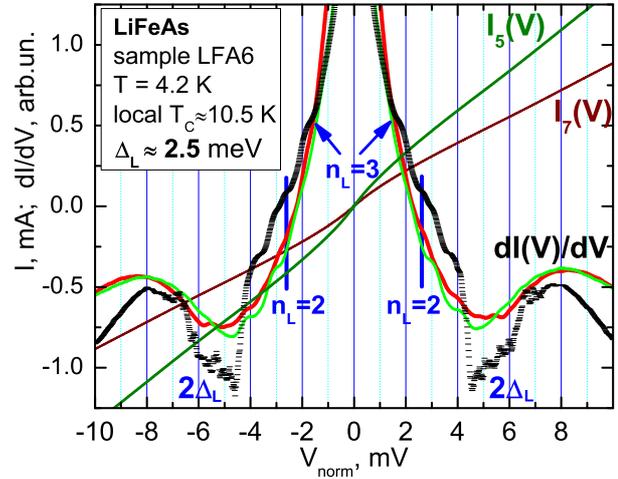}
\caption{Fig.3. Normalized CVC and dynamic conductance ($T = 4.2$ K) of SNS-Andreev stacks of 2 junctions (contacts $\sharp d5$, $\sharp d7$ plotted by light and dark solid lines, respectively) and of 3 junctions (contact $\sharp d7^{\sim}$, black dashed line). The contacts were formed by a mechanical readjustment. The well-reproduced minima ($n_L$ labels) define the $\Delta_L \approx 2.5$ meV (the $T_C^{local} \approx 10.5$ K)}
\end{figure}

The dynamic conductance spectra of contacts $\sharp d5$ and $\sharp d7$ made on LFA6 sample (see Fig.3) are presented by light and dark-grey curves. Both contacts $\sharp d5$ and $\sharp d7$ consist of two SNS-junctions in an array, while contact $\sharp d7^{\sim}$ (plotted by black dashes) is a stack of three SNS-junctions (all curves are normalized to a single junction). The smearing of the gap peculiarities (Fig.3) presumably originates from the comparatively large contact diameter ($a \approx l$). Another smearing factor might be a temperature influence on the Fermi surface, which scales with $k_BT$: for $T = 4.2$ K it yields an energy uncertainty about 0.4 meV, which is of the same order of magnitude as the value of the small gap (see Fig.4). These are, most likely, the reasons for the fuzzy small gap structure in the case. The complex structure of $n_L = 1$ minima, as well as additional peculiarities in the range of $V_1 \div V_3$ voltages, can be caused by an anisotropy of the $\Delta_L$ order parameter. By coinciding the $2\Delta_L$ peculiarities at the stack conductance curves and normalizing them to a single junction (see Fig.3), one can easily yield the large gap energy $\Delta_L \approx 2.5$ meV ($T_C^{local} \approx 10.5$ K). This value corresponds to all the contacts of Fig.3 and does not depend on the contacts resistance changing. Using our gap value, we derive a BCS-ratio for LiFeAs $2\Delta_L/k_BT_C \approx 5.5$. In an attempt to consider those conductance spectra as 4 contacts ($\sharp d5$, $\sharp d7$) and 6 contacts ($\sharp d7^{\sim}$) in an array, one could get $2\Delta_L/k_BT_C \approx 2.8$ for the large gap that is inconsistent with the BCS coupling limit of 3.52.

\begin{figure}[hc]
\includegraphics[width=.45\textwidth]{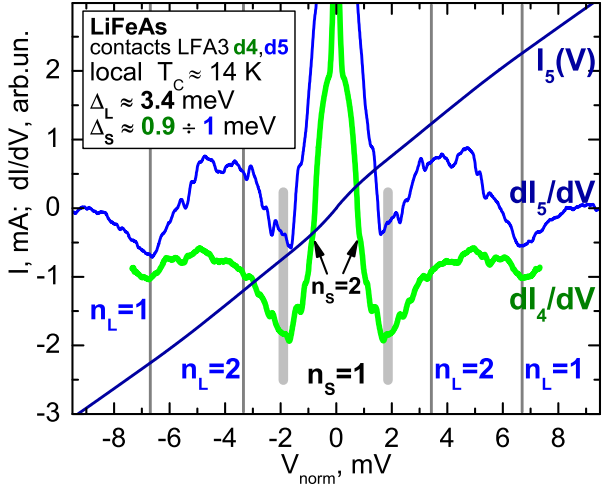}
\caption{Fig.4. Normalized to a single junction CVC and dynamic conductance ($T = 4.2$ K) of $\sharp d4$ (light curve) and $\sharp d5$ (dark curve) Andreev arrays (2 SNS-junctions in stacks). Two sets of Andreev minima ($n_{L,S}$ labels) define the large gap  $\Delta_L \approx 3.4$ meV (the $T_C^{local} \approx  14$ K) and the small gap $\Delta_S = 0.9 \div 1.0$ meV. Linear background was suppressed}
\end{figure}

The SNS-Andreev spectra of stack contacts with the maximal $T_C^{local} \approx 14$ K are shown at Fig. 4. The arrays representing two-contact stacks were formed by sequent junction readjustment. The CVC and the dynamic conductance comprising two sets of Andreev peculiarities in agreement with the formula (1) yield the values $\Delta_L \approx 3.4$ meV and $\Delta_S \approx 0.9$ meV (contact $\sharp d4$), $\Delta_S \approx 1$ meV (contact $\sharp d5$). For the contact $\sharp d5$, the $n_S = 1$ peculiarity was taken as the center of the doublet. Such a structure may be caused by a s-wave symmetry distortion of $\Delta_S$ due to the complex form of the hole-like Fermi surface sheets. The reproducibility of the small gap minima ($n_S = 1$) at $dI(V)/dV$-characteristics of Andreev stacks confirms the existence of the small superconducting gap and suggests it to be a bulk property of LiFeAs. The results presented at Figs.2,3 allow the same conclusion for the large gap. The minima $n_S = 2$ for the small gap become observable on subtracting a strait line at the $dI(V)/dV$ raise region near zero bias.

The temperature affecting on Andreev peculiarities is shown at Fig.5a. The $T_C^{local} \approx 12.5$ K is obtained from the linearization of the CVC derivative. The dynamic conductance dips series (black labels) seemingly reveal the superconducting gap $\Delta \approx 2.5$ meV. It is easy to observe that $\Delta$-minima marked as $n_L = 1$ start to close at $T \approx 6$ K. The gap temperature dependence $\Delta(T)$ (being in fact the $V_1/2$ positions plotted versus $T$) is presented at the Fig.5a inset. The aforementioned $T \approx 6$\,K appears to be approximately a half of the $T_C^{local}$. The $\Delta(T)$ was fitted by the single-band BCS-like curve and the two-band dependence (plotted in a framework of Moskalenko \cite{Moskalenko} and Suhl \cite{Suhl} equations for a superconductor with $\Delta_L/\Delta_S \approx 3.5$ and a weak interband coupling). Both the theoretical curves tend to the $T_C^{local} = 12.5$\,K, in the same time, the best fitting is achieved if the two-band curve approaching $T=0$\,K lies above the single-band one. This gives the average gap value about $2.5$\,meV. The fitting curves behavior indicates no evidence for an induced superconductivity in the condensate under consideration. Hence, this gap parameter is related to the bands with the large, ``driving'' gap $\Delta_L$ and relevant neither to small gap, nor to a surface gap.

The comparable temperature dependences of the large gap $\Delta_L(T)$ for the contacts $\sharp d5$, LFA3 sample (see the dark curve at Fig.4) and $\sharp c$, LFA5 sample are shown at Fig.5b. The experimental points were fitted by single-gap BCS-type functions. The reproducibility of the large gap behavior is obvious. Despite the difference in the $\Delta_L$ and $T_C^{local}$ values for the contacts, the corresponding BCS-ratios $2\Delta_L/k_BT_C$ are close to each other and approximately equal to 5.6. The closer examination brings out the $\Delta_L(T)$ deviations from the BCS-like curves: all the experimental dependences slightly bend, which originates from nonzero interband interaction with $\Delta_S$-bands. As for the $\Delta_S(T)$ temperature dependence represented by small circles at Fig.5b (contact $\sharp d5$, LFA3 sample), its behavior is typical for a ``driven'' gap. The similar phenomenon was observed in Mg(Al)B$_2$ superconductor earlier \cite{JETPL04,SSC12}. The difference between $\Delta_L(T)$ and $\Delta_S(T)$ courses becomes another evidence for existence of two nearly independent superconducting condensates described by the two gap values in LiFeAs.

\begin{figure}[hc]
\includegraphics[width=.45\textwidth]{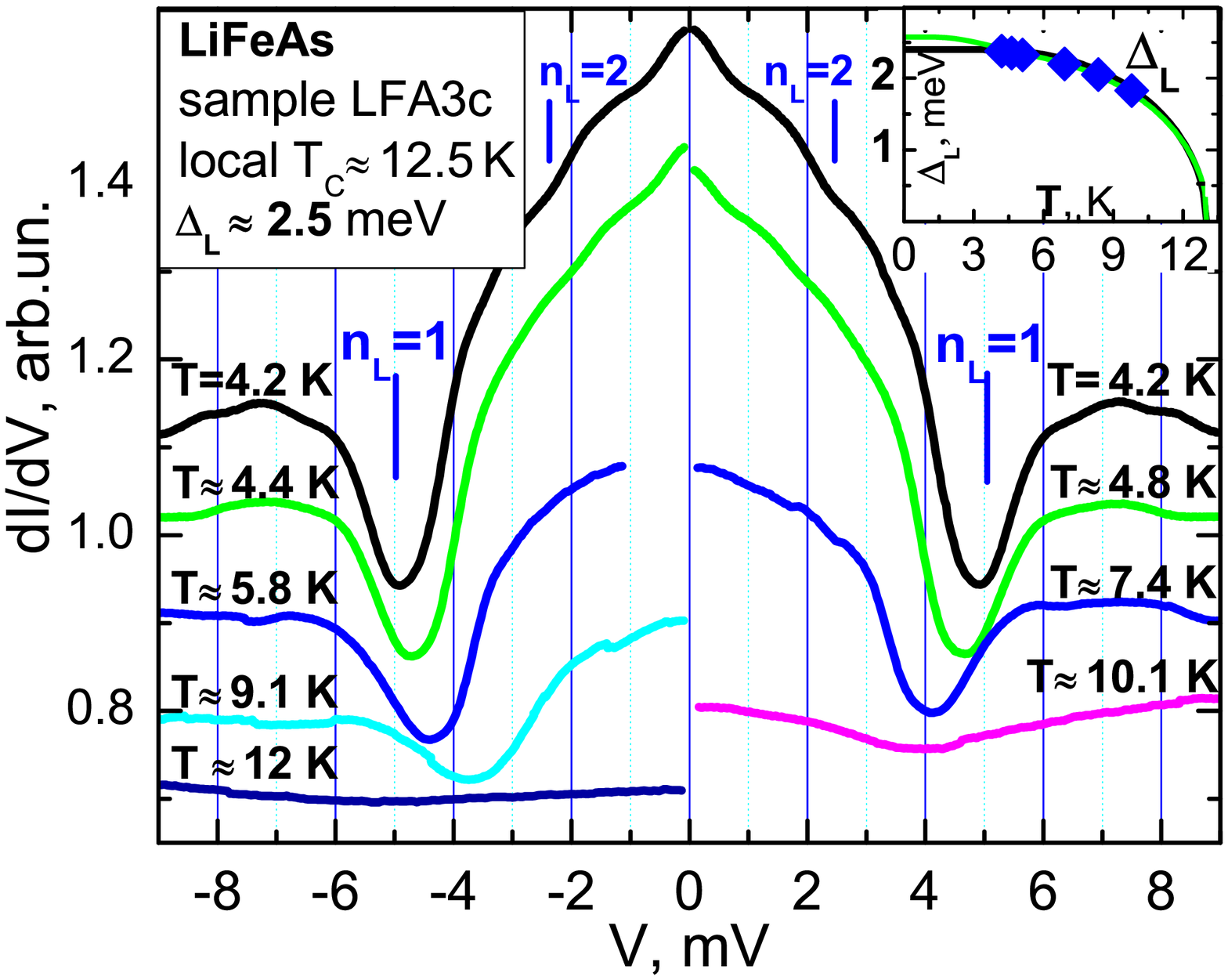}
\includegraphics[width=.45\textwidth]{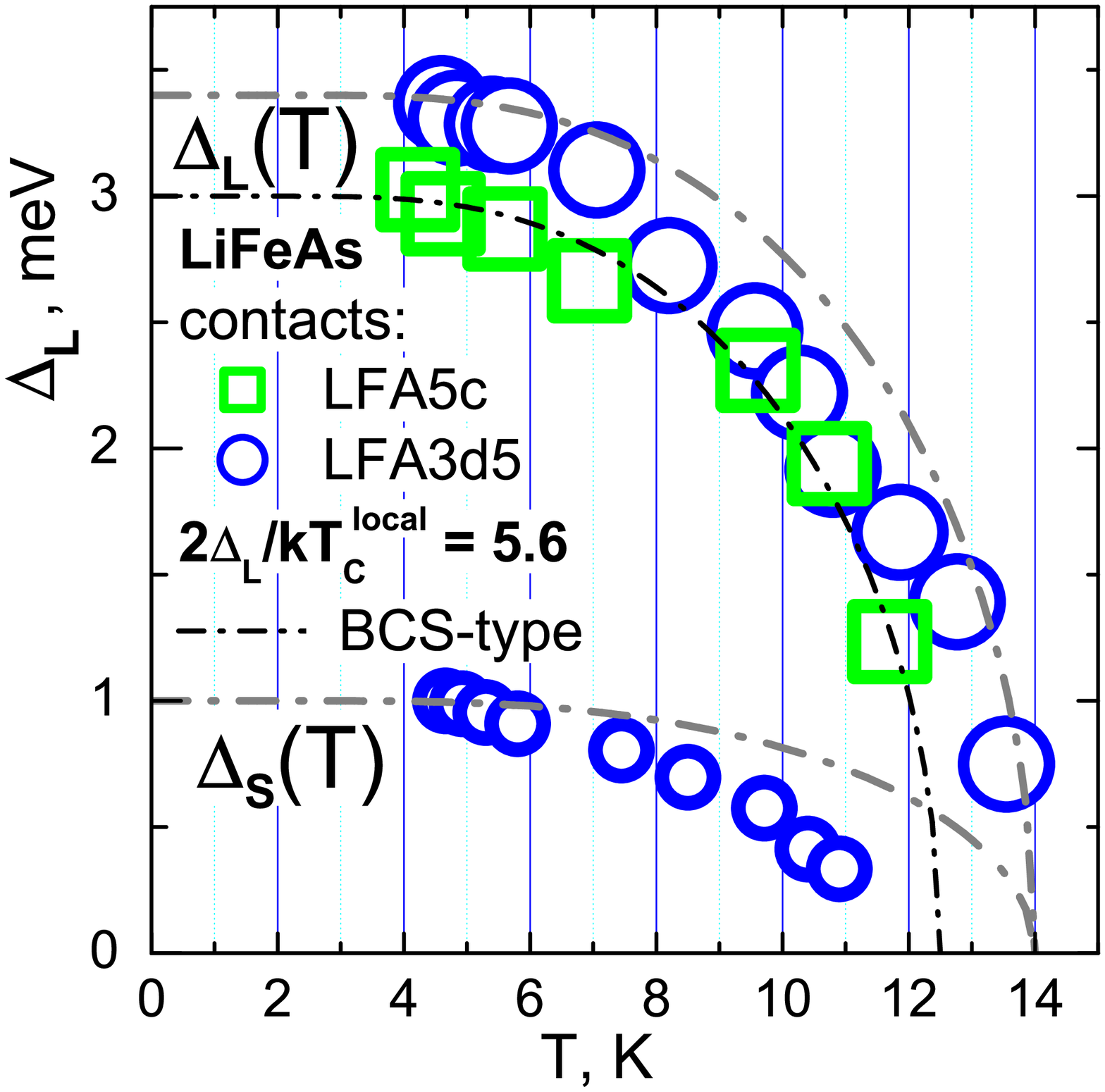}
\caption{Fig.5. \textbf{a)} Dynamic conductance of LiFeAs Andreev single SNS junction measured at the temperatures ranging from 4.2 K to $T_C^{local} \approx 12.5$ K. SGS defines the large gap $\Delta_L \approx 2.5$ meV ($n_L$ labels). The curves were shifted along the vertical scale for the sake of clarity. Inset: The large gap temperature dependence plotted using experimental data of Fig.5 (rhombs). The single-band BCS-like fit (dark line) and possible two-band fit (light line) were presented. \textbf{b)} The comparison of the large gap temperature dependences: sample LFA5, contact $\sharp c$ (squares; $T_C^{local}=12.5\pm1$\,K, $\Delta_L(4.2\,{\rm K})\approx3$\,meV), and sample LFA3, contact $\sharp d5$ (circles; $T_C^{local}=14\pm1$\,K, $\Delta_L(4.2\,{\rm K})\approx3.4$\,meV, $\Delta_S(4.2\,{\rm K})\approx1$\,meV; see Fig. 4). The point size reflects a gap value uncertainty. The single-band BCS-like dependences are plotted by dash-dotted lines}
\end{figure}

There are no published tunneling data (energetic parameters obtained with a help of SIS-, NS-, NIS-, SNS-spectroscopies) on LiFeAs so far. It was revealed in the present work for the first time. Therefore, we compare our results with the data by the number of scientific groups in the Table 1. Angle-resolved photoemission spectroscopy (ARPES) on LiFeAs single crystals \cite{Borisenko,Stockert,Inosov}, specific heat \cite{Stockert,Wei}, London penetration depth \cite{Inosov,Song,Kim}, and microwave surface impedance \cite{Imai} yield similar results for the number and values of superconducting gaps (see Table 1).
%Please note that LiFeAs single crystals from the same batch were investigated in the present work and in some studies discussed in Table 1 \cite{Borisenko,Stockert,Inosov}.
The distinction in the BCS-ratio values should be taken into account. On calculating $2\Delta/k_BT_C$ the $T_C^{onset}$ in \cite{Borisenko,Stockert,Inosov,Wei,Song,Imai,Sasmal} and $T_C^{bulk}$ in \cite{Kim} have been used, which are average values that may be higher or lower than the $T_C^{local}$ in the point of gap defining due to a samples inhomogeneity if any. By contrast, operating with the local $T_C$, one can calculate the BCS-ratio value more accurately. For the reason given above our BCS-ratio value exceeds that of other groups obtained by other methods. Nevertheless, an excellent correspondence is revealed with the $2\Delta_L/k_BT_C$ value from \cite{Sasmal} (see Table 1). For the large gap, our experimental data lead to ratio $2\Delta_L/k_BT_C = (4.6 \div 5.6)$ which exceeds the standard one 3.52 for single-gap BCS superconductors in the weak coupling limit. At the same time, the BCS-ratio for the small gap $2\Delta_S/k_BT_C \ll 3.52$ indicates the superconductivity induced by interband coupling with the "driving" bands (characterized by the large gap $\Delta_L$) in the bands with $\Delta_S$ due to k-space (internal) proximity effect \cite{Yanson} between two condensates, which resembles the situation in Mg(Al)B$_2$ \cite{SSC04,JETPL04,SSC12}, and some Fe-based superconductors \cite{LOFA,Gd,UFN,GdJPCS,FeSe}.

In conclusion, it was found that $I(V)$ and $dI(V)/dV$-characteristic of the SNS-Andreev break-junctions in LiFeAs single crystals with bulk critical temperature $Ò_Ñ = (11 \div 17)$ K could not be described in a framework of the standard single-gap model. Being observed by Andreev spectroscopy on LiFeAs for the first time, two independent subharmonic gap structures indicating the existence of multi-gap superconductivity yield the following energies of two distinct superconducting gaps:  $\Delta_L = (2.5 \div 3.4)$ meV and  $\Delta_S = (0.9 \div 1)$ meV (at $T = 4.2$ K and the $T_C^{local} \approx (10.5 \div 14)$ K). The uncertainty in the determination of the gap can be estimated to be about 10\% for large gap $\Delta_L$ and 20\% for $\Delta_S$ owing to enormous raise of dynamical conductance at low bias voltages. The estimated BCS-ratios $2\Delta_L/k_BT_C = (4.6 \div 5.6)$ exceeding the single-gap weak-coupling limit 3.52 can be caused by a strong coupling in the "driving" electron-bands \cite{Borisenko} characterized by the large gap. On the other hand, the ratio for the small gap $2\Delta_S/k_BT_C \ll 3.52$ hints on the importance of interband coupling (due to k-space proximity effect) that is to be taken into account when considering superconducting properties of LiFeAs.

The authors are grateful to Prof. Ya.G. Ponomarev
for providing techniques and materials. We also thank
S.V. Borisenko, A.N. Vasiliev and V.M. Pudalov for
helpful discussions. This work was supported by the
Russian Ministry of Education and Science (projects
11.519.11.6012, MK-3264.2012.2) and by the Russian
Foundation for Basic Research (projects 12-03-91674-ERA\_a,
12-03-01143-a). Funding by the German research society DFG
in project BE1749/13 is gratefully acknowledged.

\onecolumn
\begin{table}
 \caption{Table 1. Comparison of LiFeAs experimental data obtained by different techniques. *The mean $T_C$  defined from the $dR(T)/dT$ derivative was used. **The data range reflects an anisotropy of $\Delta_S$ order parameter \cite{Borisenko}.}
\begin{flushleft}
{\footnotesize
\begin{tabular}{|c|c|c|c|c|c|c|}
\hline
\multicolumn{7}{|c|}{\emph{Present data (SNS break-junction)}} \\
\hline
Sample  & Contact & $T_C^{local}$, K & $\Delta_L$, meV & $\Delta_S$, meV & $2\Delta_L/k_BT_C$ & $2\Delta_S/k_BT_C$\\
\hline
LFA6 & $d5, d7, d7^{\sim}, d8$ & $10.5 \pm 0.5$ & $\approx 2.5$ & - & 5.5 & - \\
\hline
LFA3b & $c$ & $12.5 \pm 0.5$ & $\approx 2.5$ & - & 4.6 & - \\
\hline
LFA3b & $d4$ & $14 \pm 0.5$ & $\approx 3.4$ & $\approx 0.9$ & 5.6 & 1.5 \\
\hline
LFA3b & $d5$ & $14 \pm 0.5$ & $\approx 3.4$ & $\approx 1$ & 5.6 & 1.6 \\
\hline
\hline
\multicolumn{7}{|c|}{\emph{Experimental data obtained on the single crystals from the same batch (as samples used by us)}} \\
\hline
Ref.  & Technique & $T_C^{onset}$, K & $\Delta_L$, meV & $\Delta_S$, meV & $2\Delta_L/k_BT_C$ & $2\Delta_S/k_BT_C$\\
\hline
Borisenko, et al. \cite{Borisenko}  & ARPES & 18 & 3.2 & $1.5 \div 2.5\,^{**}$ & $\approx 4.1$ & $1.9 \div 3.2$ \\
\hline
Stockert, et al. \cite{Stockert} & specific heat, ARPES & 17 & 2.6 & 1.2 & $\approx 3.5$ & $\approx 1.7$ \\
\hline
Inosov, et al. \cite{Inosov} & London penetration depth, ARPES & 17 & 3.1 & - & 4.1 & - \\
\hline
\multicolumn{7}{|c|}{\emph{Experimental data published by other scientific groups}} \\
\hline
Wei, et al. \cite{Wei} & specific heat & 17 & $2.7 \pm 0.8$ & $0.5 \pm 0.2$ & 3.5 & 1.2 \\
\hline
Song, et al. \cite{Song} & London penetration depth & 17.5 & 2.9 & $\approx 1.3$ & $\approx 3.8$ & $\approx 1.7$ \\
\hline
Kim, et al. \cite{Kim} & London penetration depth & $\approx 17\,^*$ & $\approx 2.8$ & $\approx 1.6$ & $\approx 3.8$ & $\approx 2.2$ \\
\hline
Imai, et al. \cite{Imai} & microwave surface impedance    & 17 & 2.9 & - & 4.0 & - \\ \cline{3-7}
 & & 16.3 & 3.0 & 1.1 & 4.2 & $\approx 1.6$ \\ \cline{3-7}
  & & 15.6 & 3.0 & 1.65 & 4.4 & $\approx 2.5$ \\
\hline
Sasmal, et al. \cite{Sasmal} & vortex penetration, reversible magnetization & $14 \div 15.3$ & 3.3 & 0.6 & 5.4 & 1 \\
\hline
\end{tabular}}
\end{flushleft}
\end{table}
\end{document}